# A Model Independent Analysis of Solar Neutrino Data

S.M. Bilenky*

*Joint Institute of Nuclear Research, Dubna, Russia*
*INFN Torino, Via P. Giuria 1, 10125 Torino, Italy*
*Dipartimento di Fisica Teorica, Università di Torino*

C. Giunti†

*INFN Torino, Via P. Giuria 1, 10125 Torino, Italy*
*Dipartimento di Fisica Teorica, Università di Torino*

(July 1994)

## Abstract

The results of a solar model independent analysis of the existing solar neutrino data under the assumption of MSW transitions in matter or vacuum oscillations (in the cases of $\nu_e$–$\nu_\mu(\nu_\tau)$ and $\nu_e$–$\nu_S$ mixing) are presented. The analysis was done in two cases. In the first case no assumptions on the value of the total fluxes of neutrinos from all reactions has been made. In the second case some wide boundaries for the values of the neutrino fluxes, that take into account the predictions of different standard solar models, are imposed. It is shown that rather large regions of the parameters $\Delta m^2$ and $\sin^2 2\theta$ are excluded by the existing data.

PACS numbers: 96.60.Kx, 14.60.Pq, 14.60.Lm.

Typeset using REVTEX

*E-mail address: BILENKY@TO.INFN.IT.

†E-mail address: GIUNTI@TO.INFN.IT.

# I. INTRODUCTION

Solar neutrino experiments are very important for the investigation of neutrino masses and mixing. Due to the large distance between the source and the detector and the small neutrino energies, solar neutrino experiments are sensitive to very small values of the difference of squared neutrino masses $\Delta m^2 = m_2^2 - m_1^2$ and (if the MSW mechanism is effective) to a wide region of mixing angles $\theta$, including the theoretically important region of small $\theta$'s.

At present there exist data of four solar neutrino experiments: the radiochemical experiments Homestake [1], Gallex [2] and Sage [3] and the direct counting experiment Kamiokande [4]. The latest experimental data are presented in Table I. The rates predicted by Bahcall and Pinsonneault (BP) [5], Turck-Chièze and Lopes (TL) [6] and Castellani, Degl'Innocenti and Fiorentini (CDF) [7] are given in the last columns of Table I. The rates predicted by these and others [8,9] Standard Solar Models (SSM's) are in agreement with each other and exceed the rates observed in all four solar neutrino experiments. One can see that the event rates observed in the Homestake and Kamiokande experiments, in which high energy neutrinos are detected, are less than, for example, the BP predicted rates by 5.5 and 3.1 standard deviations, respectively. The counting rate in the GALLEX experiment in which the small energy $pp$ neutrinos give an important contribution (about 55% according to the SSM) is about 4 standard deviations less than the predicted rate.

It was shown in Refs. [10–16] that all the existing data can be described in the framework of the SSM by the MSW mechanism [17] in the simplest case of mixing of two neutrino types. Two solutions for $\Delta m^2$ and $\sin^2 2\theta$ were found: a small mixing angle solution with $\Delta m^2 \simeq 5 \times 10^{-6}$ eV$^2$ and $\sin^2 2\theta \simeq 8 \times 10^{-3}$ and a large mixing angle solution with $\Delta m^2 \simeq 10^{-5}$ eV$^2$ and $\sin^2 2\theta \simeq 0.8$. Let us notice that the existing data can also be described by vacuum oscillations [18] if the mixing angle is large and there is a fine tuning between $\Delta m^2$ and the sun–earth distance [13].

Thus, solar neutrino data give some evidence in favour of non-zero neutrino masses and mixing if we assume that the SSM correctly predicts the fluxes of neutrinos from all neutrino producing reactions occurring in the core of the sun. However, as it is well known, the neutrino fluxes predicted by the SSM are subject to many sources of uncertainties, mainly due to a poor knowledge of some input parameters (especially nuclear cross sections and solar opacities). It is clear that any solar model independent information on the neutrino mixing parameters that can be obtained from an analysis of the data of solar neutrino experiments is very important. In this note we present the results of such a model independent analysis.

Our analysis is based on the assumption that the spectra of neutrinos from different solar reactions are known. These spectra are determined by the interactions responsible for the reactions and, as it was shown in Ref. [19], are negligibly affected by the conditions in the interior of the sun. We consider the (constant) values of the total fluxes (from $pp$, $pep$, $^7$Be, $^8$B, He$p$, $^{13}$N, $^{15}$O, $^{17}$F) as unknown parameters.

We will consider resonant MSW transitions in matter and vacuum neutrino oscillations in the case of mixing between two active neutrino states ($\nu_e$–$\nu_\mu$ or $\nu_e$–$\nu_\tau$) and mixing between $\nu_e$ and a sterile neutrino state $\nu_S$. In both cases the neutrino transition probabilities depend on two parameters: $\Delta m^2$ and $\sin^2 2\theta$. The event rate in any solar neutrino experiment is determined by the total neutrino fluxes and the values of these parameters. We use the total event rates of all four solar neutrino experiments. At fixed values of $\Delta m^2$ and $\sin^2 2\theta$ we calculate the $\chi^2$ and the corresponding confidence level for all possible values of the total neutrino fluxes. If all the confidence levels are unacceptable, the corresponding point in the $\Delta m^2$–$\sin^2 2\theta$ plane is excluded. This analysis allows us to exclude large regions of the parameter space for transitions of $\nu_e$'s into active as well as into sterile states. Let us notice that some other regions in the $\Delta m^2$–$\sin^2 2\theta$ plane were excluded in a model independent way by the Kamiokande collaboration [20] by using the data on the recoil electron spectrum in $\nu$–$e$ scattering and the data on the search for the day-night effect.

## II. THE METHOD OF ANALYSIS

The integral event rate in any experiment $a$ ($a$ = HOM (Homestake), KAM (Kamiokande), GAL (GALLEX+SAGE)[1]) is given by the expression

$$N_a = \sum_r Y_a^r \, \Phi_{\nu_e}^r \;, \tag{2.1}$$

where the index $r$ runs over all the neutrino sources which are listed in Table II. Here $\Phi_{\nu_e}^r$ is the total initial flux of $\nu_e$ from the source $r$. The initial spectrum of $\nu_e$'s from the source $r$ has the form

$$\phi_{\nu_e}^r(E) = X_{\nu_e}^r(E) \, \Phi_{\nu_e}^r \;, \tag{2.2}$$

where the functions $X_{\nu_e}^r(E)$ are normalized by the condition $\int X_{\nu_e}^r(E) \, \mathrm{d}E = 1$ and are known (see Ref. [5]). In the case of radiochemical experiments only solar $\nu_e$ on the earth are detected. We have

$$Y_a^r = \int_{E_{\mathrm{th}}^a} \sigma_a(E) \, X_{\nu_e}^r(E) \, \mathrm{P}_{\nu_e \to \nu_e}(E) \, \mathrm{d}E \;, \tag{2.3}$$

with $a$ = HOM, GAL. Here $\sigma_a(E)$ is the neutrino cross section, $\mathrm{P}_{\nu_e \to \nu_e}(E)$ is the probability of $\nu_e$ to survive and $E_{\mathrm{th}}^a$ is the threshold energy. In the case of MSW transitions the $\nu_e$ survival probability is calculated using the formula given in Ref. [21], which is valid for an exponentially decreasing electron density, and is averaged over the production region. In the case of vacuum oscillations the $\nu_e$ survival probability is averaged over the production region and over the varying distance between the sun and the earth (see Ref. [22]).

In the case of the Kamiokande experiment $\nu_e$ as well as $\nu_\mu$ (and/or $\nu_\tau$) are detected. We have

$$Y_{\mathrm{KAM}}^r = \int_{E_{\mathrm{th}}^{\mathrm{ES}}} \sigma_{\nu_e e}(E) \, X_{\nu_e}^r(E) \, \mathrm{P}_{\nu_e \to \nu_e}(E) \, \mathrm{d}E$$
$$+ \int_{E_{\mathrm{th}}^{\mathrm{ES}}} \sigma_{\nu_\mu e}(E) \, X_{\nu_e}^r(E) \, \mathrm{P}_{\nu_e \to \nu_\mu}(E) \, \mathrm{d}E \;, \tag{2.4}$$

---

[1]In our calculations we use the combined GALLEX–SAGE data: $78 \pm 10$ SNU.

where $\sigma_{\nu_\ell e}(E)$ is the cross section of the process $\nu_\ell e \to \nu_\ell e$ ($\ell = e, \mu$), $E_{\text{th}}^{\text{ES}} = \frac{1}{2}\left(T_e^{\text{th}} + \sqrt{T_e^{\text{th}}(T_e^{\text{th}} + 2m_e)}\right)$ ($T_e^{\text{th}}$ is the recoil electron kinetic energy threshold) and $P_{\nu_e \to \nu_\ell}(E)$ is the probability of transition of $\nu_e$ into $\nu_\ell$. In our calculation we took into account the efficiency and the energy resolution of the Kamiokande detector. These quantities were taken from Ref. [4].

In the analysis of solar neutrino data we take into account the luminosity constraint on the total neutrino fluxes. The thermonuclear energy of the $pp$ and CNO cycles in the sun is generated in the transition

$$4p + 2e^- \to {}^4\text{He} + 2\nu_e \,. \tag{2.5}$$

If the sun is in a quasi-static state, from Eq.(2.5) it follows that the luminosity of the sun $L_\odot$ is connected with the total neutrino fluxes by the well known relation

$$\frac{L_\odot}{4\pi d^2} = \sum_r \Phi_{\nu_e}^r \left[\frac{Q}{2} - \langle E \rangle^r\right] \,. \tag{2.6}$$

Here $d = 1\,\text{AU} = 1.496 \times 10^{13}\,\text{cm}$ is the average sun-earth distance, $L_\odot = (3.826 \pm 0.008) \times 10^{33}\,\text{erg}\,\text{sec}^{-1}$ [23], $Q = 4m_p + 2m_e - m_{{}^4\text{He}} = 26.73\,\text{MeV}$ is the total energy release in the transition (2.5) and $\langle E \rangle^r$ is the average energy of neutrinos from the source $r$. The luminosity constraint can be written down in the form of Eq.(2.1). We have

$$N_{\text{LUM}} = \sum_r Y_{\text{LUM}}^r \Phi_{\nu_e}^r \,, \tag{2.7}$$

with $N_{\text{LUM}} = L_\odot/4\pi d^2 = (8.491 \pm 0.018) \times 10^{11}\,\text{MeV}\,\text{cm}^{-2}\,\text{sec}^{-1}$ and $Y_{\text{LUM}}^r = Q/2 - \langle E \rangle^r$. The values of $\langle E \rangle^r$ and $Y_{\text{LUM}}^r$ are given in Table II.

Our procedure for the analysis of the solar neutrino data is the following. For fixed values of the parameters $\Delta m^2$ and $\sin^2 2\theta$ and fixed values of the neutrino fluxes we calculate the $\chi^2$, defined as

$$\chi^2 = \sum_a \frac{(N_a^{\text{exp}} - N_a)^2}{(\Delta N_a^{\text{exp}})^2 + \left(\frac{\Delta \sigma_a}{\sigma_a} N_a\right)^2} \,. \tag{2.8}$$

The relative uncertainties $\Delta\sigma_a/\sigma_a$ of the neutrino cross sections for the Homestake and Gallium experiments are 0.025 and 0.041, respectively (see Ref. [5]). We estimate the "goodness-of-fit" by calculating the confidence level (CL). Since we do not determine any parameter, the number of degrees of freedom of the $\chi^2$ distribution is equal to the number of data points (i.e. four: three neutrino rates and the solar luminosity). If all the confidence levels found for a given value of $\Delta m^2$, $\sin^2 2\theta$ and all possible values of the neutrino fluxes[2] are smaller than $\alpha$ (we choose $\alpha = 0.1, 0.05, 0.01$), it means that the corresponding point in the $\Delta m^2$-$\sin^2 2\theta$ plane is excluded at $100(1-\alpha)\%$ CL. In this way we obtain the exclusion plots presented in Figs.1–8. Let us notice that for the purpose of determination of the excluded regions in the parameter space our approach is the most conservative: any decrease of the number of degrees of freedom would increase the excluded regions.

---

[2] It is possible to perform this calculation by trying all possible values of the neutrino fluxes. However, in practice it is computationally much more convenient to find the minimum value of the $\chi^2$ with respect to the neutrino fluxes. We find this minimum with the Minuit program, which is available in the CERN program library. Let us emphasize that the minimization of the $\chi^2$ is only a computational trick which allows us to calculate the maximum value of the CL without having to try all the possible values of the neutrino fluxes with a very lengthy montecarlo.

For the exclusion plots presented in Figs.3, 4, 7, 8 the only requirement was that all the total neutrino fluxes are positive. Let us call this case A. However, we found that at some values of the parameters $\Delta m^2$ and $\sin^2 2\theta$ the minimum of $\chi^2$ is achieved with unreasonable values of some neutrino fluxes (e.g. 100 times larger than the corresponding SSM fluxes). Thus we also considered the following case B: the different solar neutrino fluxes are allowed to vary in the interval $\xi^r_{\min} \Phi^r_{\nu_e}(BP) \leq \Phi^r_{\nu_e} \leq \xi^r_{\max} \Phi^r_{\nu_e}(BP)$, where $\Phi^r_{\nu_e}(BP)$ are the BP–SSM value of the neutrino fluxes and the factors $\xi^r_{\min}$ and $\xi^r_{\max}$ are chosen in such a way to include the predictions of the different existing solar models [5–9,24]. The values of these factors are given in Table II. We determined the minimum (maximum) values for the $pp$, $pep$, $^7$Be and He$p$ fluxes by subtracting (adding) 3 times the range of solar model predictions to the minimum (maximum) predicted flux (notices that this range is larger than the $1\sigma$ error given by BP). Since it has been recently suggested [25] that the value of the astrophysical factor $S_{17}(0)$ could be significantly lower than that used in SSM calculations, we let the $^8$B flux to be arbitrarily small. In the recent Dar and Shaviv calculation [24] the CNO fluxes are very small. In order to cover this possibility we let also the CNO fluxes to be arbitrarily small. We determined the maximum values of the $^8$B and CNO fluxes by adding 3 times the $1\sigma$ error calculated by BP to the BP average value.

The excluded regions of the parameters $\Delta m^2$ and $\sin^2 2\theta$ in case B are presented in Figs.5 and 6. As it can be seen from a comparison of these figures with Figs.3 and 4 the excluded regions in case B are larger than in the case where no limitation is imposed on the values of the neutrino fluxes.

Let us emphasize that the limits for the solar neutrino fluxes in case B are chosen arbitrarily. Nevertheless, it is interesting and instructive to investigate how the forbidden regions in the $\Delta m^2$–$\sin^2 2\theta$ plane change if some limits on the allowed values of the neutrino fluxes are imposed. The case B illustrates these changes. Let us also stress that the limits on the allowed values of the neutrino fluxes which we imposed in case B are rather wide in comparison with the predictions of the existing solar models.

### III. RESULTS AND DISCUSSION

#### A. Vacuum Oscillations (Constant Transition Probability)

We discuss here the results of a model independent analysis of solar neutrino data under the assumption that vacuum neutrino oscillations take place and that the oscillating terms in the transition probability vanish due to averaging over the neutrino spectrum, the region of the sun where neutrinos are produced, and so on. In this case the transition probabilities are constant and are determined by the elements of the mixing matrix. In the case of oscillations between $n$ neutrino states the minimum value of the averaged $\nu_e$ survival probability is equal to $\overline{P}^{\min}_{\nu_e \to \nu_e} = 1/n$ (see Ref. [18]). This minimum corresponds to maximum neutrino mixing. We consider transitions of solar $\nu_e$ into active as well as into sterile states[3].

The curves in Fig.1 represent the maximum confidence levels of the fits as functions of the survival probability $\overline{P}_{\nu_e \to \nu_e}$ for transitions of $\nu_e$ into $\nu_\mu$ and/or $\nu_\tau$. It can be seen from Fig.1 that the following values of $\overline{P}_{\nu_e \to \nu_e}$ are excluded at 90% CL:

$$\overline{P}_{\nu_e \to \nu_e} < 0.11 \quad \text{and} \quad 0.18 < \overline{P}_{\nu_e \to \nu_e} < 0.45 \quad (A) ,$$
$$\overline{P}_{\nu_e \to \nu_e} < 0.54 \quad \text{and} \quad \overline{P}_{\nu_e \to \nu_e} > 0.89 \quad (B) .$$

The following values of $\overline{P}_{\nu_e \to \nu_e}$ are excluded at 95% CL:

$$\overline{P}_{\nu_e \to \nu_e} < 0.09 \quad (A) ,$$
$$\overline{P}_{\nu_e \to \nu_e} < 0.43 \quad \text{and} \quad \overline{P}_{\nu_e \to \nu_e} > 0.95 \quad (B) .$$

From Fig.1 one can also see that $\overline{P}_{\nu_e \to \nu_e} = 1/2$ and $\overline{P}_{\nu_e \to \nu_e} = 1/3$, which correspond to a maximum mixing in the case of two and three oscillating neutrino states, respectively, are excluded at 87% and 93% CL in case A and at 92% and 98% CL in case B.

In Fig.2 the maximum confidence level as a function of the survival probability $\overline{P}_{\nu_e \to \nu_e}$ is presented for the case of transitions of solar $\nu_e$'s into sterile states. From Fig.2 one can see that the following values are excluded at 90% CL:

$$\overline{P}_{\nu_e \to \nu_e} < 0.61 \quad (A) ,$$
$$\overline{P}_{\nu_e \to \nu_e} < 0.66 \quad \text{and} \quad \overline{P}_{\nu_e \to \nu_e} > 0.86 \quad (B) .$$

From Fig.2 one can also see that $\overline{P}_{\nu_e \to \nu_e} = 1/2$ is excluded at 97% and at 98% CL in cases A and B, respectively.

### B. MSW Transitions in Matter

Here we discuss the results of a model independent analysis of the solar neutrino data in the case of resonant MSW transitions of solar $\nu_e$'s into $\nu_\mu$ (or $\nu_\tau$) or sterile neutrinos. Let us remind that a MSW resonance takes place if at some point $x_R$ the following equation is satisfied

$$2\sqrt{2}\, G_{\rm F}\, E\, N_e(x_R) = \Delta m^2 \cos 2\theta . \qquad (3.1)$$

Here $G_{\rm F}$ is the Fermi constant, $E$ is the neutrino energy and $N_e$ is the electron density. The electron density has a maximum in the center of the sun: $N_e^c \simeq 100 N_{\rm A}$ cm$^{-3}$, where $N_{\rm A}$ is the Avogadro number. From Eq.(3.1) it follows that MSW transitions can take place if the following condition is satisfied:

---

[3]The difference between these two types of transitions is due to the Kamiokande data: $\nu_\mu$ and/or $\nu_\tau$ give a contribution to $\nu$–$e$ scatterings detected in the Kamiokande experiment while sterile neutrinos $\nu_{\rm S}$ do not interact with matter.

$$E \gtrsim \frac{\Delta m^2 \cos 2\theta / \text{eV}^2}{1.5 \times 10^{-5}} \text{ MeV} . \qquad (3.2)$$

Let us consider first MSW transitions of $\nu_e$'s into $\nu_\mu$ (or $\nu_\tau$). The region in the $\Delta m^2$–$\sin^2 2\theta$ plane excluded by the existing solar neutrino data in case A (no limits on the values of the neutrino fluxes) is presented in Fig.3. In this figure we have also plotted (shaded areas) the two allowed regions (90% CL) found by us from an analysis of the data with the BP neutrino fluxes. These allowed regions practically coincide with those found by other authors [10–16]. The excluded region has the triangular shape typical of a strong $\nu_e$ suppression. Since this region extends below $\Delta m^2 \cos 2\theta \simeq 3 \times 10^{-6}$ eV$^2$, from Eq.(3.2) it is clear that in this region a MSW resonance takes place for neutrinos from all sources. In this region the value of the flux of $pp$ $\nu_e$'s on the earth is strongly suppressed. This suppression of the flux of $pp$ $\nu_e$'s is in contradiction with the Gallium data. It cannot be counterbalanced by a large initial $pp$ flux $\Phi_{\nu_e}^{pp}$ because $\Phi_{\nu_e}^{pp}$ is limited by the luminosity constraint. The observed Gallium event rate cannot be due to a high value of $\Phi_{\nu_e}^{^8\text{B}}$ because the $^8$B neutrino flux is constrained by the data of the Kamiokande experiment in which both $\nu_e$ and $\nu_\mu$ are detected (notice that in most of the excluded region the $^8$B $\nu_e$ flux on the earth is not suppressed because the corresponding MSW transition is highly non–adiabatic). The observed Gallium event rate cannot either be due to high values of the other neutrino fluxes ($pep$, $^7$Be and CNO) because the values of these fluxes are constrained by the data of the Homestake experiment.

The result of the calculation of the excluded regions in the $\Delta m^2$–$\sin^2 2\theta$ plane for $\nu_e$–$\nu_S$ MSW transitions in case A is presented in Fig.4. The allowed region (90% CL) found with the BP neutrino fluxes is also shown in Fig.4 (shaded area). This allowed region practically coincides with that found in Ref. [12]. It can be seen from Fig.4 that there are two excluded regions of parameters in the case of transitions of $\nu_e$'s into sterile neutrinos. The large region is similar to the excluded region in the case of $\nu_e \to \nu_\mu(\nu_\tau)$ transitions. The other region is specific for $\nu_e \to \nu_S$ transitions. Since in this region $\Delta m^2 \gtrsim 5 \times 10^{-5}$ eV$^2$, from Eq.(3.2) it follows that this region corresponds to a MSW suppression of the flux of high energy $^8$B $\nu_e$'s. A strong suppression of high energy $^8$B $\nu_e$'s only is not allowed by the data of the solar neutrino experiments. In fact, this suppression cannot be counterbalanced by a large initial total $^8$B flux (to fit the Kamiokande data) because it would give too large Clorine and Gallium event rates (in contradiction with the Homestake and GALLEX+SAGE data) due to the unsuppressed low energy part of the $^8$B neutrino flux.

Now we discuss what happens with the exclusion regions if limits on the solar neutrino fluxes are imposed (case B). The excluded regions for MSW $\nu_e \to \nu_\mu(\nu_\tau)$ and $\nu_e \to \nu_S$ transitions in case B are shown in Figs.5 and 6. A comparison of Fig.3 and Fig.5 (Fig.4 and Fig.6) shows that the region of values of the parameters $\Delta m^2$ and $\sin^2 2\theta$ that is forbidden by the existing solar neutrino data is strongly increased if we put some limits on the possible values of the neutrino fluxes. Figs.5 and 6 illustrate the fact that even assuming rather wide limits for the values of the solar neutrino fluxes, the existing solar neutrino data strongly restrict the region of possible values of the parameters $\Delta m^2$ and $\sin^2 2\theta$. The triangular excluded region for $\Delta m^2 \lesssim 5 \times 10^{-6}$ eV$^2$ is due to a strong suppression of the flux of low energy $pp$ $\nu_e$'s. The excluded region with $5 \times 10^{-6}$ eV$^2 \lesssim \Delta m^2 \lesssim 10^{-4}$ eV$^2$ is due to a large suppression of the flux of $^8$B $\nu_e$'s. As it can be seen from Figs.5 and 6, this excluded region is larger in the case of $\nu_e \to \nu_S$ transitions than in the case of $\nu_e \to \nu_\mu(\nu_\tau)$ transitions. This is connected with the fact that the sterile neutrinos that are produced in $\nu_e \to \nu_S$ transitions do not contribute to the Kamiokande event rate, while the $\nu_\mu(\nu_\tau)$ that are produced in $\nu_e \to \nu_\mu(\nu_\tau)$ transitions interact with electrons (with a cross section that is about 0.15 of the cross section for $\nu_e$–$e$ scattering). Let us notice that in this region, corresponding to a large suppression of $^8$B $\nu_e$'s, it is also impossible to fit the Clorine and Gallium data: the large flux of $^7$Be neutrinos that is necessary to fit the Homestake data would give an excessive Gallium event rate. In the large (left) part of the excluded region $P_{\nu_e \to \nu_e} = 1$, which is excluded at 98% CL.

### C. Vacuum Oscillations

In this section we consider vacuum neutrino oscillations in the case in which the oscillating term in the transition probability does not vanish due to averaging. As it is well known, in the case of oscillations between two neutrino states ($\nu_e \to \nu_\mu(\nu_\tau)$ or $\nu_e \to \nu_S$) the $\nu_e$ survival probability is given by

$$P_{\nu_e \to \nu_e}(E, R) = 1 - \frac{1}{2} \sin^2 2\theta \left(1 - \cos \frac{\Delta m^2 R}{2E}\right), \quad (3.3)$$

where $R$ is the distance between the point where neutrinos are produced and the earth. In the expression for the observable event rates this probability is integrated over the neutrino energy, over the region of the sun where neutrinos are produced and over the varying distance between the sun and the earth. If the average neutrino oscillation length is comparable with the sun–earth distance and if the mixing angle $\theta$ is large, then the cosine term in Eq.(3.3) could give a non–zero contribution to the event rates. The regions of values of the parameters $\Delta m^2$, $\sin^2 2\theta$ which allow to describe the experimental data (with the solar neutrino fluxes predicted by the BP–SSM) were given in Ref. [13]. We have found the regions of values of the parameters $\Delta m^2$ and $\sin^2 2\theta$ that are excluded for any value of the initial neutrino fluxes (case A).

For the problem of the "survival" of the contribution of the cosine term in Eq.(3.3) the most important integration is the one over energy. Let us write down this term as

$$\cos \pi \frac{x}{x_0}, \quad (3.4)$$

where $x = 1/E$ and $x_0 = 2\pi/\Delta m^2 R$. Let $\overline{x}$ be the characteristic length over which the product of the neutrino cross section and the neutrino flux changes. If $x_0 \ll \overline{x}$ the cosine term does not give any contribution to the integral. If $x_0 \gg \overline{x}$ the neutrino oscillations are not observable ($P_{\nu_e \to \nu_e} = 1$). Only in the case $x_0 \simeq \overline{x}$ we can expect that the cosine term gives a non–zero contribution to the event rate. In this case, for the parameter $\Delta m^2$ we have the following estimate:

$$\Delta m^2 \simeq \frac{2\pi \overline{E}}{d} \simeq 8.3 \times 10^{-12} \,\text{eV}^2 \, \frac{\overline{E}}{\text{MeV}}, \quad (3.5)$$

where $\overline{E}$ is the average neutrino energy. For $pp$, $^7$Be and $^8$B neutrinos we have $\Delta m^2 \simeq 2.5 \times 10^{-12}$ eV$^2$, $\Delta m^2 \simeq 7.5 \times 10^{-12}$ eV$^2$, $\Delta m^2 \simeq 5.8 \times 10^{-11}$ eV$^2$, respectively.

Fig.7 shows the regions of the values of the parameters $\Delta m^2$ and $\sin^2 2\theta$ that are excluded in a model independent way by the existing experimental data in the case of $\nu_e \to \nu_\mu(\nu_\tau)$ vacuum oscillations. The allowed regions (90% CL) of the values of the parameters that was obtained by us with the BP-SSM values of the neutrino fluxes are also shown in Fig.7 (shaded areas). These regions practically coincide with the allowed regions found in Ref. [13]. In the large excluded region with $\overline{\Delta m^2} \simeq 2.5 \times 10^{-12}$ eV$^2$ the flux of $pp$ neutrinos is strongly suppressed. As we discussed above, this suppression is in contradiction with the data of the Gallium experiments (if the constraints from the other experiments and from the solar luminosity are taken into account). The other excluded regions with $\Delta m^2 < 5 \times 10^{-11}$ eV$^2$ are due to an excessive suppression of the fluxes of $pp$ and $^7$Be neutrinos.

The excluded regions of the values of the parameters $\Delta m^2$ and $\sin^2 2\theta$ in the case of $\nu_e \to \nu_S$ vacuum oscillations are shown in Fig.8. As in the case of MSW transitions in matter, the excluded regions due to a strong suppression of the high energy $^8$B neutrinos are larger in the case of $\nu_e \to \nu_S$ oscillations than in the case of $\nu_e \to \nu_\mu(\nu_\tau)$ oscillations. In this case we did not find any allowed region with the SSM-BP neutrino fluxes (in agreement with Ref. [13]).

## IV. CONCLUSION

We have presented here the results of a model independent analysis of the existing solar neutrino data. We have considered resonant MSW transitions in matter and vacuum oscillations in the case of mixing between two neutrino types ($\nu_e$–$\nu_\mu(\nu_\tau)$ and $\nu_e$–$\nu_S$, where $\nu_S$ is a sterile neutrino). We have obtained *forbidden* regions of the values of the parameters $\Delta m^2$ and $\sin^2 2\theta$ without any assumption about the values of the total neutrino fluxes from all sources (taking into account only the luminosity constraint). We have also obtained the forbidden regions in the $\Delta m^2$–$\sin^2 2\theta$ plane in a case in which some wide limits, that take into account the predictions of the different solar models, are imposed on the values of the total neutrino fluxes. We have shown that in this model independent approach the existing solar neutrino data allow to forbid rather large regions of values of the parameters $\Delta m^2$ and $\sin^2 2\theta$ (expecially when limits on the values of the total neutrino fluxes are imposed).

## ACKNOWLEDGMENTS

It is a pleasure for us to express our gratitude to G. Conforto, S. Degl'Innocenti, P. Krastev, F. Martelli, S. Petcov and A. Smirnov for very useful discussions. We would like to thank B. Balantekin and W. Haxton for the hospitality at Institute for Nuclear Theory of the University of Washington where part of this work was carried out.

# REFERENCES

[1] R. Davis Jr., Talk presented at the 6$^{th}$ International Workshop on Neutrino Telescopes, Venezia, March 1994.

[2] GALLEX Collaboration, Phys. Lett. B **285**, 376 (1992); Phys. Lett. B **314**, 445 (1993); Phys. Lett. B **327**, 377 (1994).

[3] V.N. Gavrin, Talk presented at the 6$^{th}$ International Workshop on Neutrino Telescopes, Venezia, March 1994.

[4] K. S. Hirata et al., Phys. Rev. Lett. **65**, 1297 (1990); Phys. Rev. D **44**, 2241 (1991); Y. Suzuki, Talk presented at the 6$^{th}$ International Workshop on Neutrino Telescopes, Venezia, March 1994.

[5] J.N. Bahcall and R. Ulrich, Rev. Mod. Phys. **60**, 297 (1988); J.N. Bahcall, *Neutrino Physics and Astrophysics*, Cambridge University Press, 1989; J.N. Bahcall and M.H. Pinsonneault, Rev. Mod. Phys. **64**, 885 (1992).

[6] S. Turck-Chièze, S. Cahen, M. Cassé and C. Doom, Astrophys. J. **335**, 415 (1988); S. Turck-Chièze and I. Lopes, Astrophys. J. **408**, 347 (1993); S. Turck-Chièze et al., Phys. Rep. **230**, 57 (1993).

[7] V. Castellani, S. Degl'Innocenti and G. Fiorentini, Astronomy & Astrophysics **271**, 601 (1993); S. Degl'Innocenti, INFN-FE-07-93.

[8] I.J. Sakmann, A.I. Boothroyd and W.A. Fowler, Astrophys. J. **360**, 727 (1990).

[9] G. Berthomieu et al., Astronomy & Astrophysics **268**, 775 (1993).

[10] GALLEX Collaboration, Phys. Lett. B **285**, 390 (1992).

[11] X. Shi, D.N. Schramm and J.N. Bahcall, Phys. Rev. Lett. **69**, 717 (1992).

[12] S.A. Bludman, N. Hata, D.C. Kennedy and P.G. Langacker, Phys. Rev. D **47**, 2220 (1993); N. Hata and P.G. Langacker, Phys. Rev. D **50**, 632 (1994).

[13] P.I. Krastev and S.T. Petcov, Phys. Lett. B **299**, 99 (1993); Phys. Rev. Lett. **72**, 1960 (1994).

[14] L.M. Krauss, E. Gates and M. White, Phys. Lett. B **299**, 94 (1993); Fermilab-Pub-94/176-A (hep-ph@xxx.lanl.gov/9406396).

[15] G.L. Fogli and E. Lisi, Astropart. Phys. **2**, 91 (1994).

[16] G. Fiorentini et al., Phys. Rev. D **49**, 6298 (1994).

[17] L. Wolfenstein, Phys. Rev. D 17 (1978) 2369; S. P. Mikheyev and A. Y. Smirnov, Il Nuovo Cimento C 9 (1986) 17.

[18] S.M. Bilenky and B. Pontecorvo, Phys. Rep. 41 (1978) 225.

[19] J.N. Bahcall, Phys. Rev. D **44**, 1644 (1991).

[20] K. S. Hirata et al., Phys. Rev. Lett. **65**, 1301 (1990); Phys. Rev. Lett. **66**, 9 (1991).

[21] S.T. Petcov, Phys. Lett. B **200**, 373 (1988); P.I. Krastev and S.T. Petcov, Phys. Lett. B **207**, 64 (1988); T.K. Kuo and J. Pantaleone, Phys. Rev. D **39**, 1930 (1989); Rev. Mod. Phys. **61**, 937 (1989).

[22] P.I. Krastev and S.T. Petcov, Phys. Lett. B **285**, 85 (1992).

[23] Review of Particle Properties, Phys. Rev. D **45**, Part II (1992).

[24] A. Dar and G. Shaviv, Technion-Ph-94-5 (astro-ph/9401043).

[25] T. Motobayashi et al., Yale-40609-1141.

TABLES

TABLE I. Data of solar neutrino experiments and rates predicted by BP [5], TL [6] and CDF [7]. $N_{\text{KAM}}^{\text{BP}}$ is the Kamiokande event rate predicted by BP.

| Experiment | Event Rate (SNU) | SSM Predictions (SNU) | | |
|---|---|---|---|---|
| | | BP | TL | CDF |
| Kamiokande | $N_{\text{KAM}}^{\text{exp}}/N_{\text{KAM}}^{\text{BP}} = 0.51 \pm 0.04 \pm 0.06$ | $1 \pm 0.14$ | $0.8 \pm 0.2$ | 0.98 |
| Homestake | $N_{\text{HOM}}^{\text{exp}} = 2.32 \pm 0.23$ | $8.0 \pm 1.0$ | $6.4 \pm 1.4$ | 7.8 |
| GALLEX | $N_{\text{GAL}}^{\text{exp}} = 79 \pm 10 \pm 6$ | $131.5_{-6}^{+7}$ | $123 \pm 7$ | 131 |
| SAGE | $N_{\text{GAL}}^{\text{exp}} = 74 \pm 19 \pm 10$ | | | |

TABLE II. Solar neutrino fluxes (with $1\sigma$ errors) predicted by BP. $\langle E \rangle$ is the average neutrino energy, $Y_{\text{LUM}} = Q/2 - \langle E \rangle$, where $Q = 26.73\,\text{MeV}$, and $\xi_{\text{min}}$ and $\xi_{\text{max}}$ determine the limits for the values of the total neutrino fluxes in case B.

| Reaction | $\langle E \rangle$ (MeV) | $Y_{\text{LUM}}$ (MeV) | $\Phi_{\nu_e}$(BP) (cm$^{-2}$sec$^{-1}$) | $\xi_{\text{min}}$ | $\xi_{\text{max}}$ |
|---|---|---|---|---|---|
| $pp$ | 0.265 | 13.10 | $(6.00 \pm 0.004) \times 10^{10}$ | 0.93 | 1.07 |
| $pep$ | 1.442 | 11.92 | $(1.43 \pm 0.02) \times 10^{8}$ | 0.61 | 1.29 |
| $^7$Be | 0.813 | 12.55 | $(4.89 \pm 0.29) \times 10^{9}$ | 0.46 | 1.40 |
| $^8$B | 6.710 | 6.66 | $(5.69 \pm 0.82) \times 10^{6}$ | 0 | 1.43 |
| He$p$ | 9.625 | 3.74 | $1.23 \times 10^{3}$ | 0.90 | 1.13 |
| $^{13}$N | 0.7067 | 12.66 | $(4.92 \pm 0.84) \times 10^{8}$ | 0 | 1.51 |
| $^{15}$O | 0.9965 | 12.37 | $(4.26 \pm 0.82) \times 10^{8}$ | 0 | 1.58 |
| $^{17}$F | 0.9994 | 12.37 | $(5.39 \pm 0.86) \times 10^{6}$ | 0 | 1.48 |

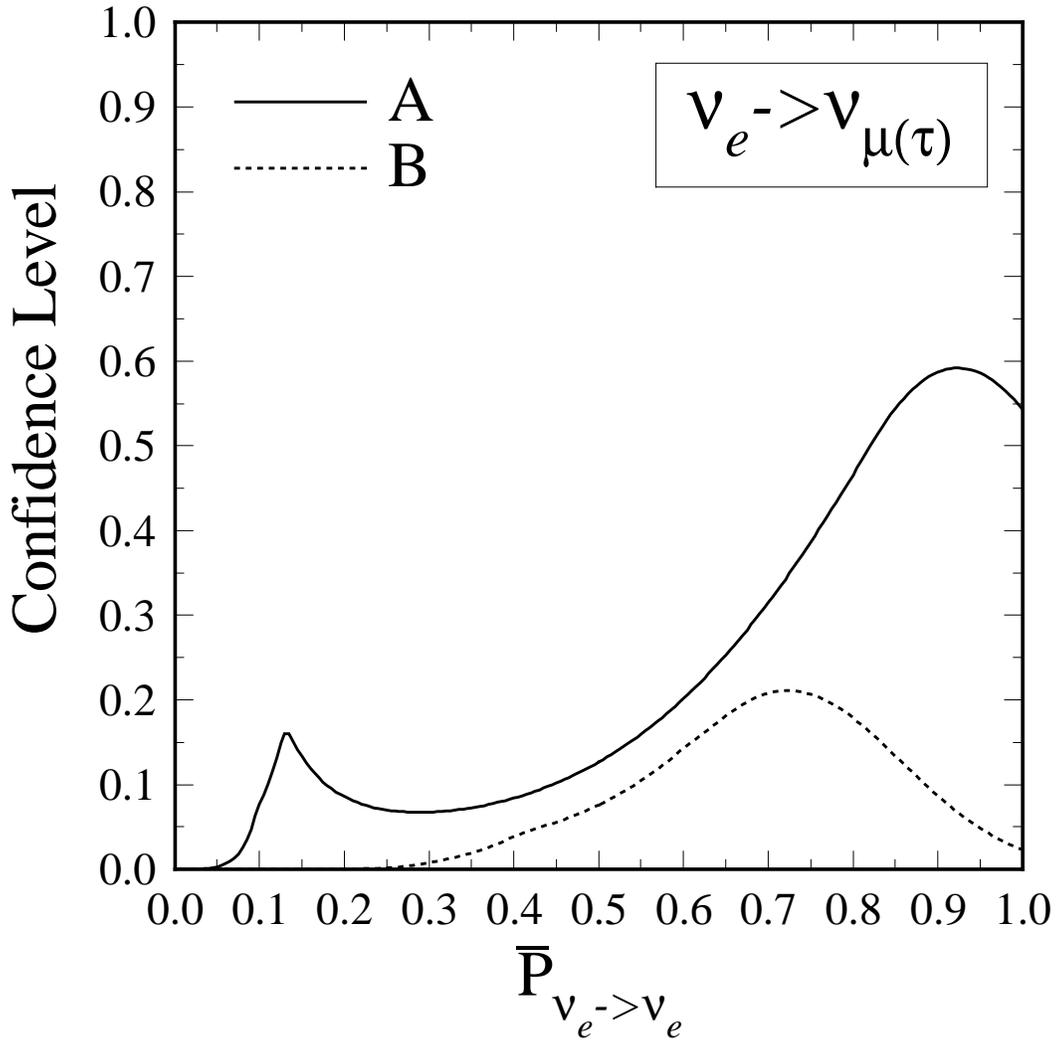

FIG. 1. Maximum Confidence Level as a function of the survival probability $\overline{P}_{\nu_e \to \nu_e}$ in the case of a constant averaged probability for transitions of solar $\nu_e$'s into $\nu_\mu$ and/or $\nu_\tau$. Curve A was obtained in case A, in which no limits on the values of the neutrino fluxes are assumed. Curve B corresponds to case B, in which some limits on the values of the neutrino fluxes are imposed.

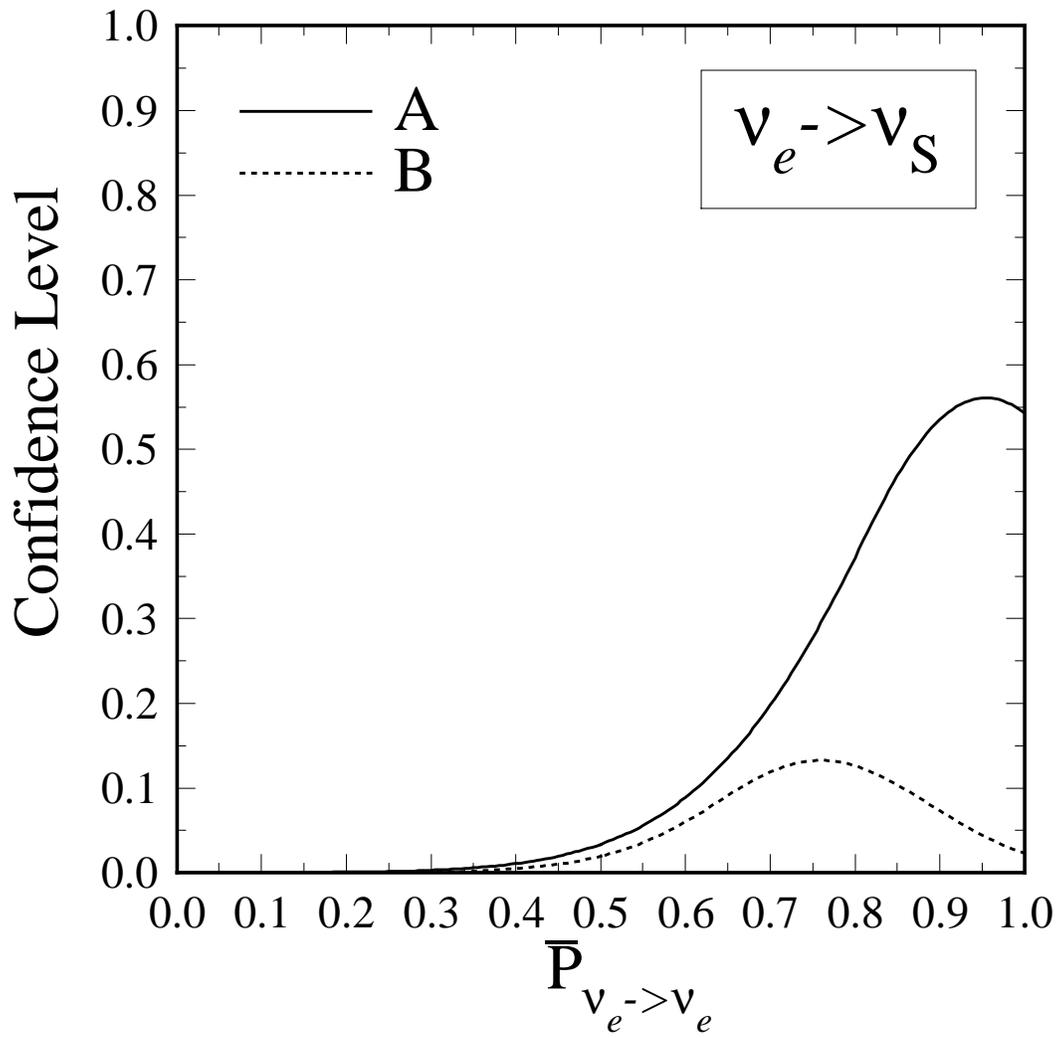

FIG. 2. Maximum Confidence Level as a function of the survival probability $\overline{P}_{\nu_e \to \nu_e}$ in the case of a constant averaged probability for transitions of solar $\nu_e$'s into sterile neutrinos. Curve A was obtained in case A, in which no limits on the values of the neutrino fluxes are assumed. Curve B corresponds to case B, in which some limits on the values of the neutrino fluxes are imposed.

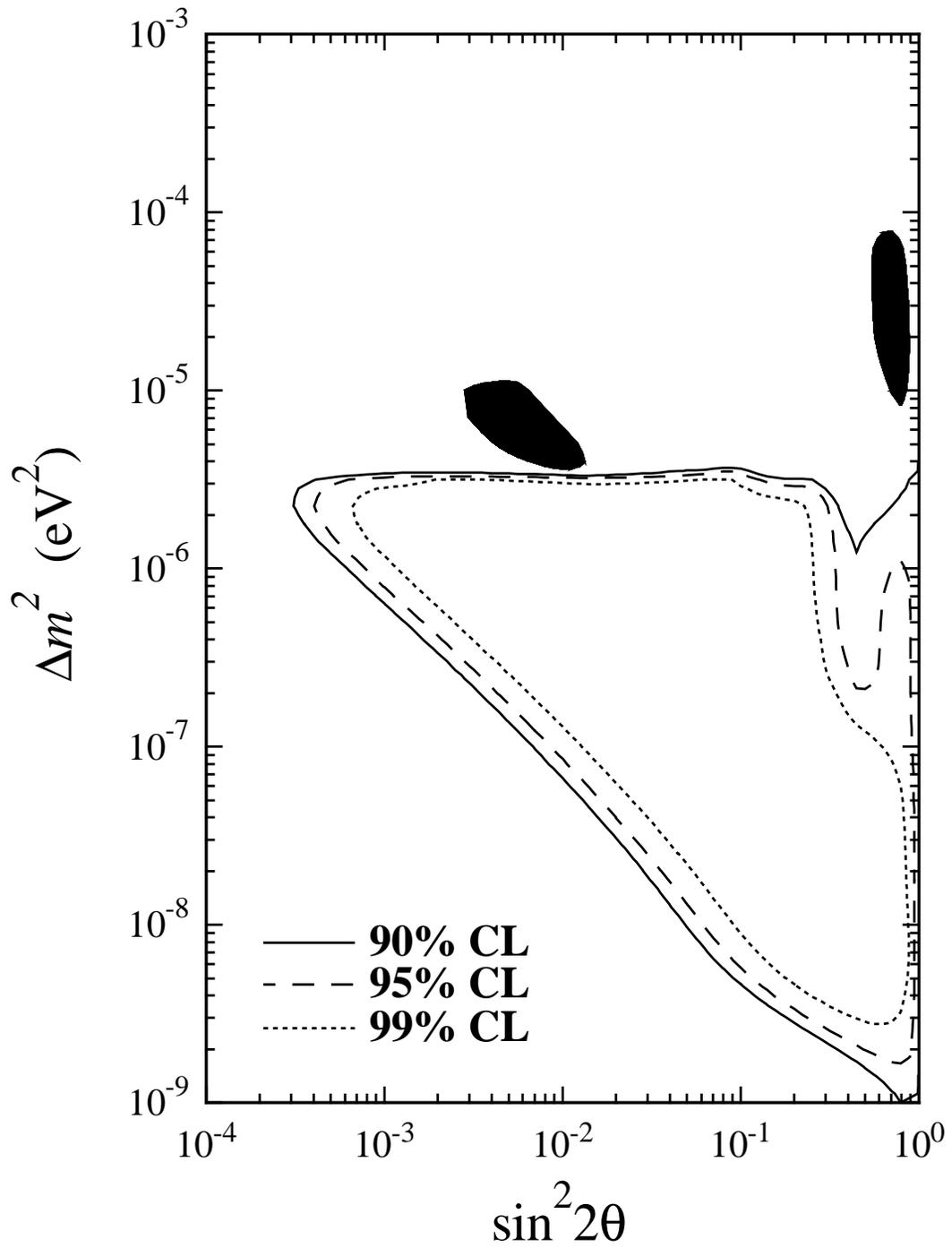

FIG. 3. MSW transitions: excluded regions in the $\sin^2 2\vartheta$–$\Delta m^2$ plane for $\nu_e$–$\nu_\mu(\nu_\tau)$ mixing in case A (no limits on the values of the neutrino fluxes are assumed). The allowed regions found with the BP neutrino fluxes are also shown (shaded areas).

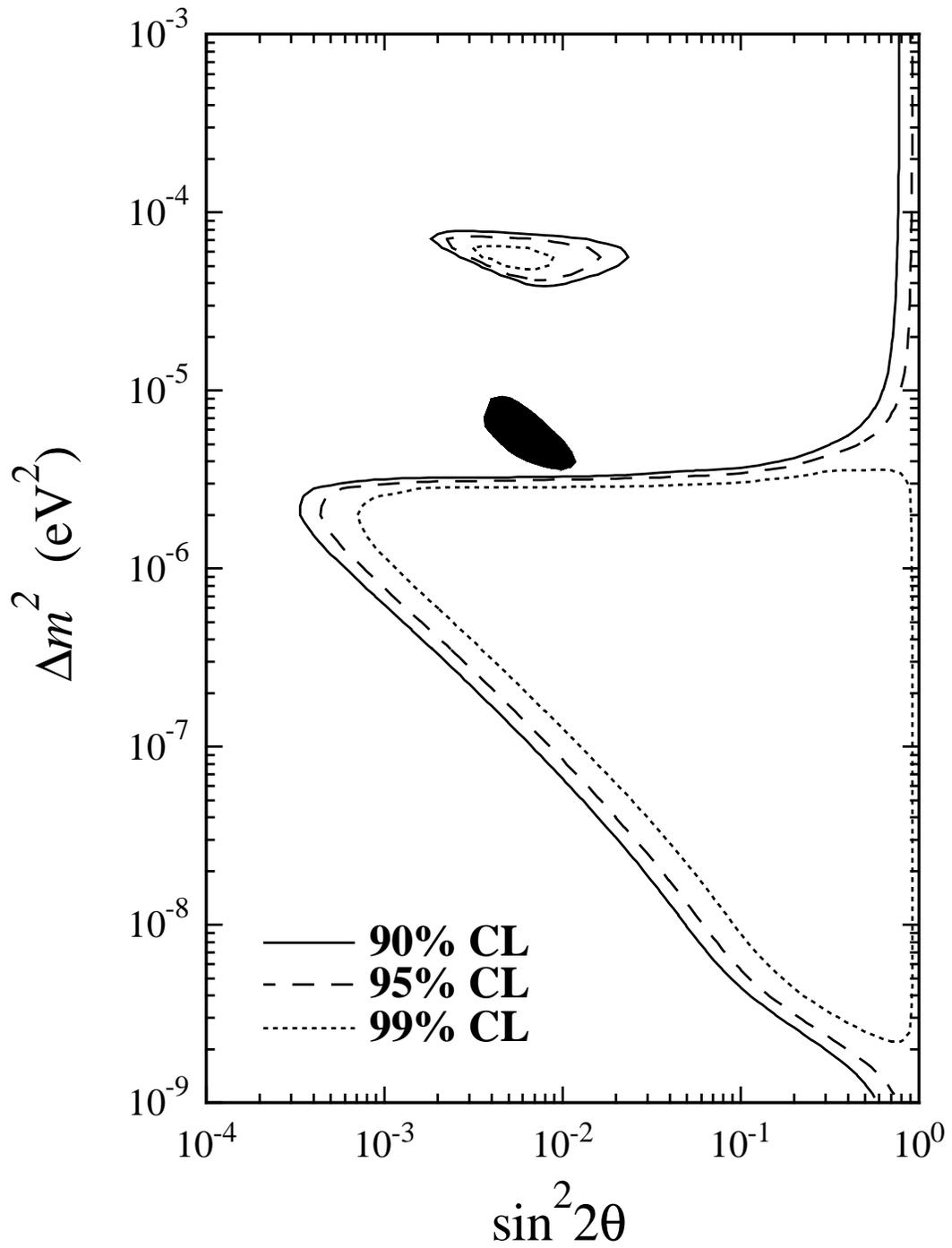

FIG. 4. MSW transitions: excluded regions in the $\sin^2 2\vartheta$–$\Delta m^2$ plane for $\nu_e$–$\nu_S$ mixing in case A (no limits on the values of the neutrino fluxes are assumed). The allowed region found with the BP neutrino fluxes is also shown (shaded area).

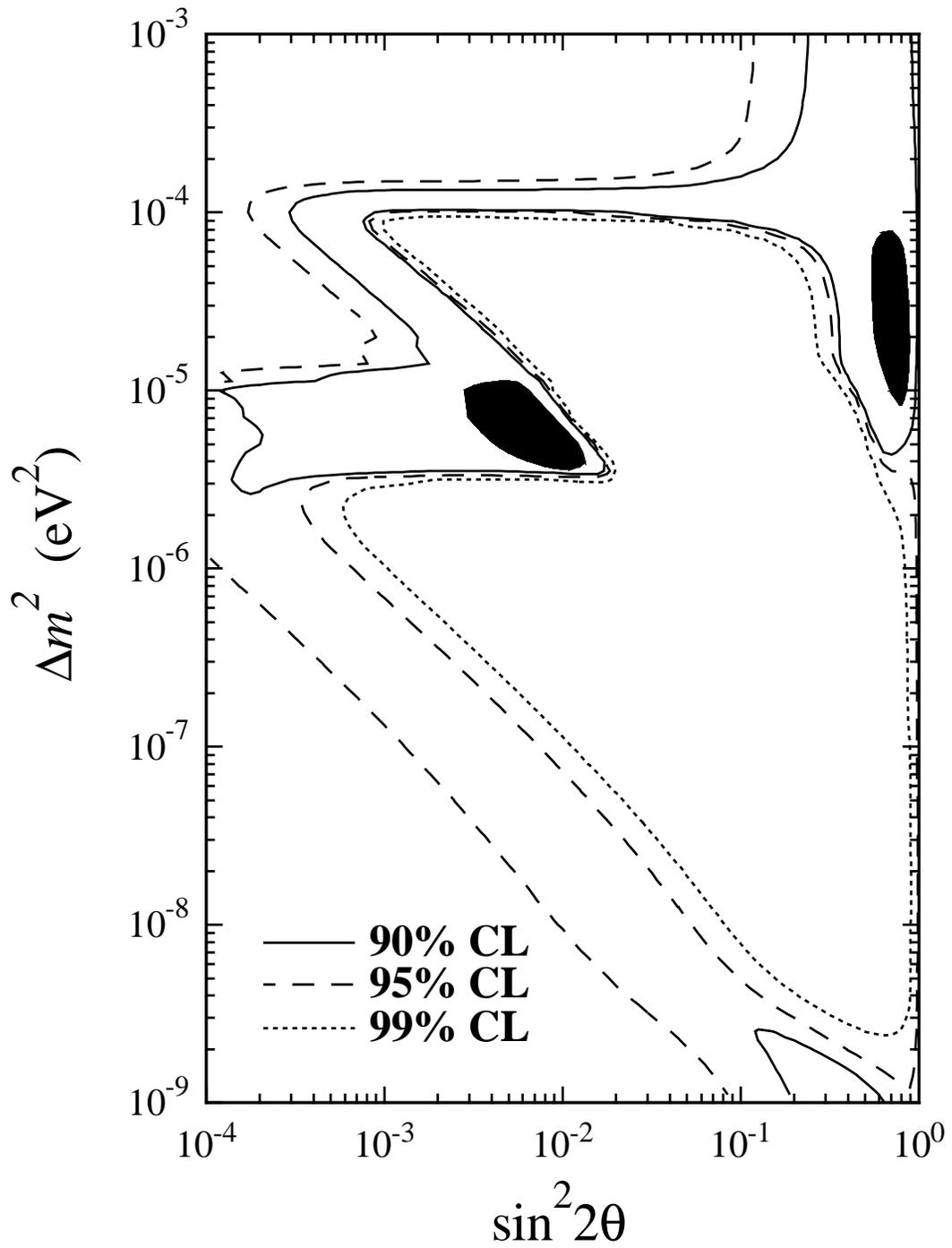

FIG. 5. MSW transitions: excluded regions in the $\sin^2 2\vartheta$–$\Delta m^2$ plane for $\nu_e$–$\nu_\mu(\nu_\tau)$ mixing in case B (some limits on the values of the neutrino fluxes are imposed). The allowed regions found with the BP neutrino fluxes are also shown (shaded areas).

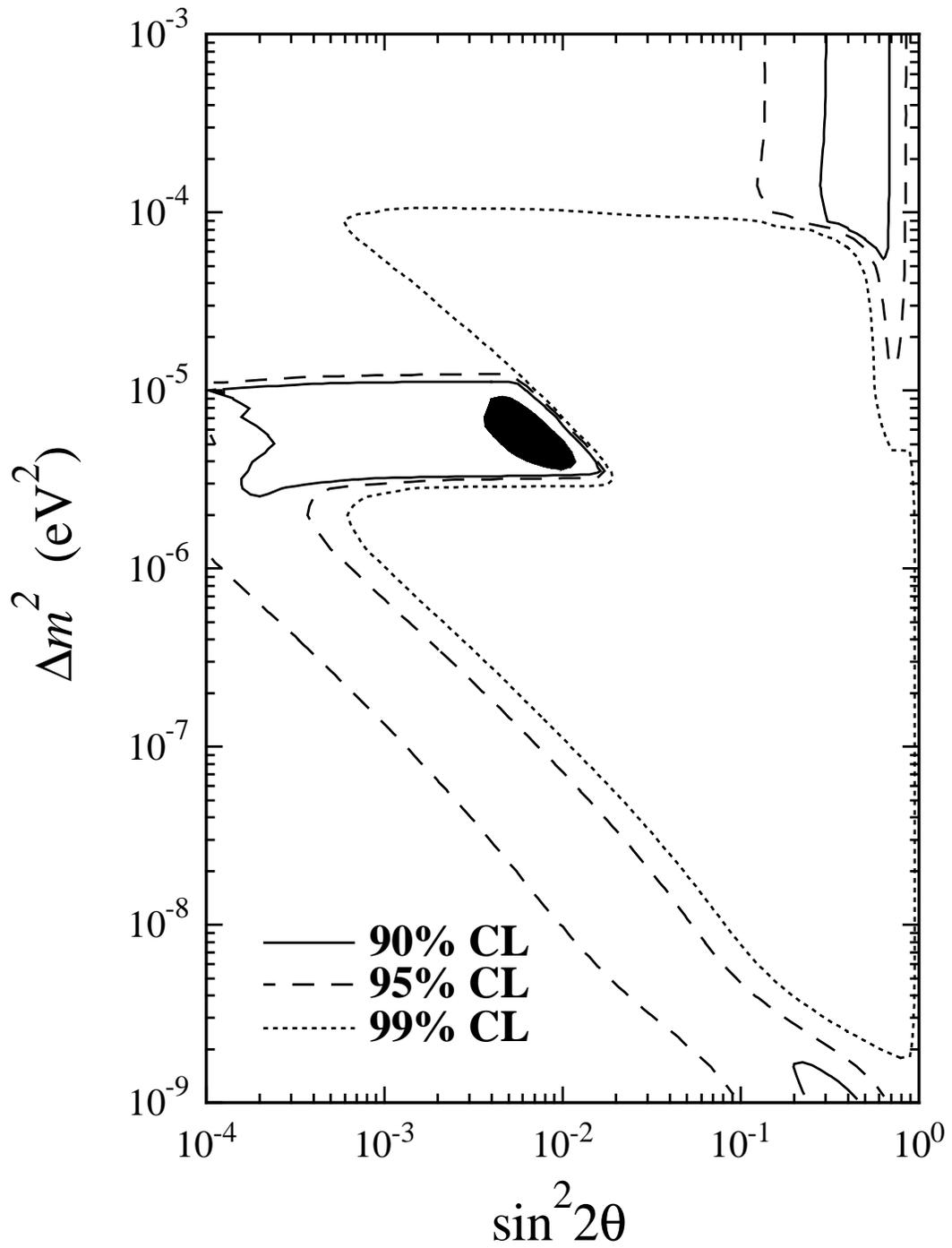

FIG. 6. MSW transitions: excluded region in the $\sin^2 2\vartheta$–$\Delta m^2$ plane for $\nu_e$–$\nu_S$ mixing in case B (some limits on the values of the neutrino fluxes are imposed). The allowed region found with the BP neutrino fluxes is also shown (shaded area).

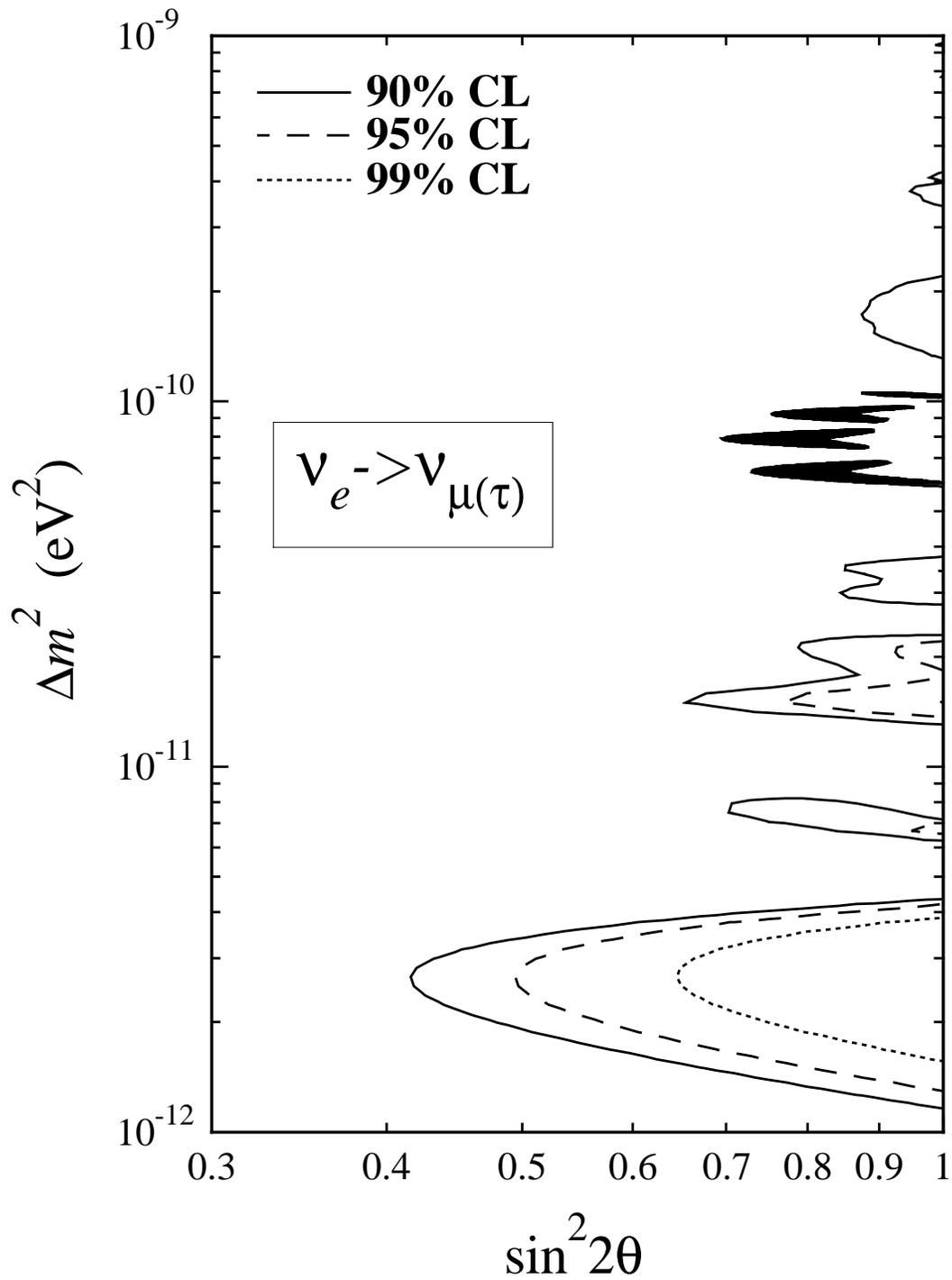

FIG. 7. Vacuum oscillations: excluded regions in the $\sin^2 2\vartheta$–$\Delta m^2$ plane for $\nu_e$–$\nu_\mu(\nu_\tau)$ mixing in case A (no limits on the values of the neutrino fluxes are assumed). The allowed regions found with the BP neutrino fluxes are also shown (shaded areas).

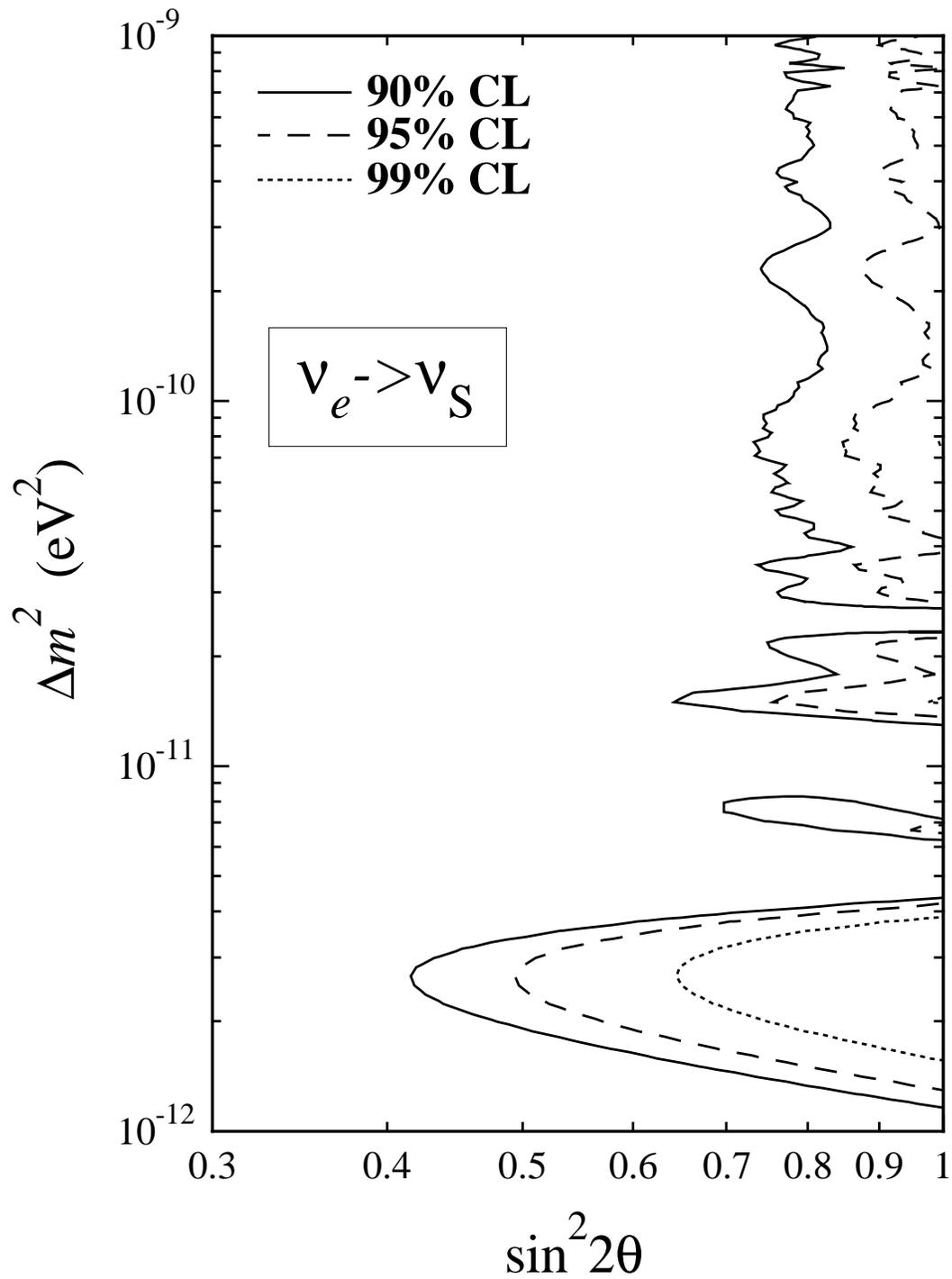

FIG. 8. Vacuum oscillations: excluded regions in the $\sin^2 2\vartheta$–$\Delta m^2$ plane for $\nu_e$–$\nu_S$ mixing in case A (no limits on the values of the neutrino fluxes are assumed).